\documentclass[prl,twocolumn,amsfonts,amssymb,amsmath,floatfix,tightenlines,superscriptaddress,longbibliography ]{revtex4-1}

\usepackage{amsfonts,amssymb}
\usepackage{amsmath,times}
\usepackage{graphicx}
\usepackage{hyperref}    
\usepackage{color}
\usepackage{cleveref}
\usepackage[lofdepth,lotdepth,caption=false]{subfig}
\usepackage[normalem]{ulem}

\begin{document}

\title{Universal coherent atom-molecule oscillations in the dynamics
of the unitary Bose gas near a narrow Feshbach resonance}

\author{Ke Wang}
\affiliation{Department of Physics and James Franck Institute, University of Chicago, Chicago, Illinois 60637, USA}
\affiliation{Kadanoff Center for Theoretical Physics, University of Chicago, Chicago, Illinois 60637, USA}
  \author{Zhendong Zhang}
  \affiliation{E. L. Ginzton Laboratory and Department of Applied Physics, Stanford University, Stanford, CA 94305, USA}
\author{Shu Nagata}
\affiliation{Department of Physics and James Franck Institute, University of Chicago, Chicago, Illinois 60637, USA}
   \author{Zhiqiang Wang}
\affiliation{Department of Physics and James Franck Institute, University of Chicago, Chicago, Illinois 60637, USA}
\affiliation{Hefei National Research Center for Physical Sciences at the Microscale and School of Physical Sciences, University of Science and Technology of China, Hefei, Anhui 230026, China}

\affiliation{Shanghai Research Center for Quantum Science and CAS Center for Excellence in Quantum Information and Quantum Physics, University of Science and Technology of China, Shanghai 201315, China}

\affiliation{Hefei National Laboratory, University of Science and Technology of China, Hefei 230088, China
}

\author{K. Levin}
\affiliation{Department of Physics and James Franck Institute, University of Chicago, Chicago, Illinois 60637, USA}

\begin{abstract}
Quench experiments on a unitary Bose gas around a broad Feshbach
resonance have led to the discovery of universal dynamics. This
universality is manifested in the measured atomic momentum distributions
where, asymptotically, a quasi-equilibrated metastable state is found in
which both the momentum distribution and the time scales are determined
by the particle density. In this paper we present counterpart studies
but for the case of a very narrow Feshbach resonance of $^{133}$Cs atoms
having a width of 8.3 mG. In dramatic contrast to the behavior reported
earlier, a rapid quench of an atomic condensate to unitarity is observed
to ultimately lead to coherent oscillations involving dynamically
produced condensed and non-condensed molecules and atoms. The same
characteristic frequency, determined by the Feshbach coupling, is
observed in all types of particles. To understand these quench dynamics
and how these different particle species are created, we develop a beyond Hartree-Fock-Bogoliubov dynamical framework including a new type of cross correlation between atoms and molecules. This leads to a
quantitative consistency with the measured frequency. Our results, which
can be applied to the general class of bosonic superfluids associated
with narrow Feshbach resonances, establish a new paradigm for universal
dynamics dominated by quantum many-body interactions.
\end{abstract}

\maketitle

{\it Introduction.} Understanding the inherently unstable unitary Bose gas has remained a challenge
~\cite{PhysRevLett.101.135301,PhysRevLett.107.135301,PhysRevLett.102.090402,Piatecki2014,PhysRevLett.108.145305,PhysRevLett.106.153005}.
Some progress, however, has been made principally because rapid field sweeps 
across a Feshbach resonance to unitarity show, fortuitously, that the gas
lives long enough to reveal features of quasi-steady state behavior before inevitable losses set in. This is seen through the momentum distribution
$n(k)$ of non-condensed particles
~\cite{Makotyn2014,Hadzibabic}, 
{  which are associated with nonzero momentum.}
Importantly, these sweeps establish that the signal at high momentum
$k$ grows and eventually saturates, as a function of time.
These intriguing saturation phenomena have been referred to as a form of prethermalization~\cite{PhysRevA94.039901,Ranccon2013}.
Interestingly, for the systems studied thus far, the time scale for such quasi-steady state dynamics is claimed to be universal. 
As the
scattering length
$a_s \rightarrow \infty$, it is presumed that
time dependent phenomena should be determined by a single energy scale $E_F
 \propto
( 6 \pi^2 n)^{2/3}$~\cite{PhysRevLett.88.210403,PhysRevLett.103.025302,
PhysRevA.81.063613,PhysRevA.84.033618,PhysRevLett.108.195301,PhysRevA.89.033614,Gao2020},
where
$n$ is the density of bosons.

Here we
emphasize that this form of universality does not apply to the narrow
resonance case.
Even at unitarity, 
the relevant energy scale corresponds to the many-body Feshbach coupling $\alpha \sqrt{n}$, which is 
to be contrasted with the Fermi energy $E_{\mathrm{F}}$.
This scale may be ignored in previous work~\cite{Makotyn2014,Hadzibabic} since there $\alpha\sqrt{n}\gg E_F $.
However,
the
dynamics of the unitary gas in this narrow resonance
regime which has not received the same degree of attention,
remains to be experimentally characterized and theoretically understood.

\begin{figure}
\includegraphics[width=80mm,clip]
{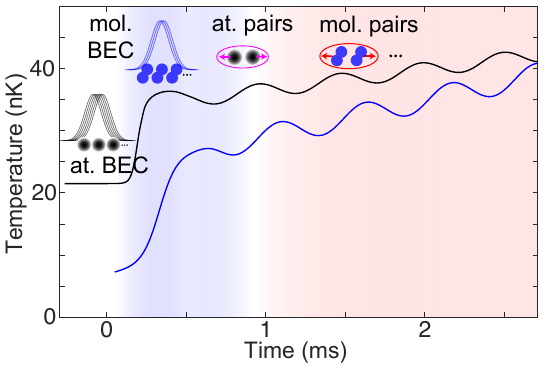}
\caption{Effective temperature of atoms (black solid line) and molecules (blue) based on an error-bar weighted fit to data
from Ref.\cite{Zhang2023}
emphasizing the
creation of 
out-of-condensate particles deriving from an atomic BEC
after a quench to unitarity, see Supplement~\cite{supp}. The small cartoons indicate the
different species of particles generated at different stages of the dynamics, including atomic and molecular BECs, and finite-momentum atomic and molecular pairs. 
}
\label{fig1}
\end{figure}

In this paper, using both theory and experiment, we investigate the
case of this narrow Feshbach resonance.
In contrast to previous Bose gas literature which focused on the atoms,
one has access to a sizeable population of
closed channel molecules which govern much of the physics
at unitarity. 
Our work
is motivated by an earlier observation of the effective temperature, $T_{\rm{eff}}$ 
of out-of-condensate atoms and of closed channel
molecules~\cite{Zhang2023} which is plotted in
Fig.~\ref{fig1}.
These are created during a quench to unitarity of
an atomic Bose-Einstein condensate (BEC) 
of $^{133}$Cs atoms associated with a narrow resonance (of width 8.3 mG).
Although previously oscillations in $T_{\rm{eff}}$ were not reported, they are revealed  
here based on an error-bar weighted fit to the temperature dependent data.

It is our goal to experimentally investigate and
theoretically understand the initial creation and subsequent oscillatory dynamics of these 
excited atomic
and molecular states.
To this end, 
we employ
a time-of-flight (TOF) procedure.
By analyzing a data set from previous experiments~\cite{Zhang2023,wang2023stability}, we are able
to characterize multiple new observables. These include
the atomic and molecular particle number distribution at different momenta and their associated
kinetic energies~\cite{supp}.

Theoretically to understand the oscillatory behavior 
observed in these quantities, we go beyond the  Hartree-Fock-Bogolibov (HFB)
~\cite{Kokkelmans2002,Snyder2012}
approach 
and establish a new dynamical framework 
that includes atom-molecule correlations. These correlations, which have no counterpart in the broad resonance case, turn out to 
be central to the dynamics
of the out-of-condensate particles. 
In the process we identify three distinct evolutionary stages in the quench dynamics;
these are initiated by the generation of a molecular BEC
and subsequently followed by the creation of finite-momentum atoms and molecules which
then undergo coherent oscillations\cite{Zhang2023,tian2024,Sadhasivam:2024usq}.
Satisfactory quantitative agreement is obtained between theory and
experiment both for the 
rather large oscillatory frequency
which turns out to be around
$2$ kHz
and for the net increase of kinetic energy 
(associated with the decay of the initial atomic condensate) 
which is transferred to
newly 
excited atoms and molecules.

{\it Dynamical Framework. }
The effective Hamiltonian for a magnetic Feshbach resonance
is associated with coupling between atomic and molecular channels (respectively called ``open" and ``closed"). It is given by
\begin{eqnarray}
\label{2}
&& \hat H=  \sum_\sigma \int d^3 x \hat{\psi}^\dagger_\sigma(x) \left( -\frac{\hbar^2\nabla^2}{2m_\sigma}+\nu_\sigma \right)  \hat{\psi}_\sigma(x)  \\
 && +   \int d^3 x    \sum_\sigma \frac{g_\sigma}{2} \hat\psi^{\dagger}_\sigma  \hat\psi^{\dagger}_\sigma   \hat\psi_\sigma \hat\psi_\sigma
    - \left(\alpha  \hat{\psi}^\dagger_1  \hat{\psi}^\dagger_1 \hat{\psi}_2 +h.c.  \right).\nonumber
\end{eqnarray}
Here $\sigma=1,2$ represents the atoms and molecules,
respectively, and $\nu_\sigma=\delta_{\sigma,2}\nu$ where $\nu$ is the detuning; $g_\sigma$ is proportional to the background $s$-wave interaction, and $\alpha$ corresponds to
the important Feshbach coupling \cite{Duine2004} strength between the two channels.
Model parameters are related to the physical quantities by $g_\sigma=4\pi \hbar^2 a_{\sigma} /(\Gamma_\sigma m_\sigma)$, 
$\alpha=\Gamma_1^{-1} \sqrt{2\pi \hbar^2 a_1 \delta \mu \Delta B/m_1 }$ and $\Gamma_\sigma$ is $1-2  a_{\sigma} \Lambda/\pi$.
  Here $a_1$ and $a_2$ are the background scattering length of atom and molecules, $\delta \mu$ is the relative magnetic moment between open and closed channels and $\Delta B$ is the resonance width.   The detuning $\nu$ is related to the physical detuning $\nu_r=\delta \mu \delta B$  by $\nu=\nu_r+ \Lambda m_1 \Gamma_1 \alpha^2  /(\pi^2\hbar^2) $ where $\delta B$ is the distance of the magnetic field from the resonance. {  Here \(\Lambda\) is a regularization cutoff, chosen here as \(a_1 \Lambda = \pi/10\), although the results are not sensitive to the specific numerical value.} We use experimental values~\cite{Zhang2023} for the parameters
$a_1,a_2,\delta \mu, \Delta B$ to determine $g_1, g_2$ and $\alpha$, and the initial atomic density $n$ in simulations~\cite{supp}.

To theoretically study non-equilibrium dynamics, we focus on one and two-point equal-time correlations as dynamical variables
\cite{Phys.Rev.A.102.063314,Phys.Rev.A98.053612,PhysRevResearch.6.L012056}.
The one-point correlation $\xi_\sigma(t)=V^{-1/2}\langle \hat{\psi}_\sigma(k=0,t )\rangle$ represents the atomic
or molecular condensate wavefunction, and $c_\sigma\equiv |\xi_\sigma(t)|^2$ represents the condensate density. 
Operators representing the excitations are then given by $\hat \psi_\sigma'(k)=\hat \psi_\sigma(k)-\delta_{k,0} \sqrt{V} \xi_\sigma(t)$. 

It is convenient
to introduce a 4-vector field operator, $\hat\Psi(k)~=[\hat\psi'_1(k), \hat\psi'^\dagger_{1}(-k), \hat\psi'_{2}(k), \hat\psi'^\dagger_{2}(-k) ]$. The quantum fluctuations that involve the finite momentum particles correspond to
two-point correlation functions, $G_{\alpha \beta}(k)=\langle \hat\Psi_{\alpha}(-k) \hat\Psi_{\beta}(k) \rangle$, from which we deduce physical quantities. The variables $n_1(k) \equiv  G_{21}(k)$ and $n_2(k)\equiv G_{43}(k)$ represent the particle number 
associated with finite-$k$ atoms and molecules. Important are the interspecies correlations $ G_{13}(k)$ 
and $ G_{23}(k)$ which are central to the dynamics: $G_{13}(k)$ is associated with the molecular pairs and $G_{23}(k)$ controls the particle flow which results in the creation of finite-$k$ molecules.

{ The dynamical equations associated with these correlation functions
are readily derived 
in terms of coupling constants $g_1$, $g_2$ and $\alpha$ which are, in turn, defined in terms of physical
scattering lengths and the resonance width associated with the Hamiltonian of Eq.~\ref{2}.} These equations are given by
\begin{eqnarray}
&&i  \partial_t \xi_{1}= 2g_1  n_1\xi_{1} +\left(g_1\xi_1^2+g_1 x_1-2\alpha \xi_2  \right)\xi^*_1-2\alpha f_{12},  
\\
&&i  \partial_t \xi_{2} = \left( \nu+2g_2  n_2\right)\xi_{2} + (g_2\xi_2^2+g_2 x_2) \xi^*_{2}- \alpha (x_1+\xi_1^2)
.\nonumber
\label{4}
\end{eqnarray}
Here $n_{\sigma}=\sum_{k\neq 0} n_\sigma(k)/V$ is the density of out-of-condensate atoms/molecules, {  which includes all finite-$k$ population of atoms/molecules},  and $V$ is the volume. 
We define $x_1=\sum_{k\neq 0} G_{11}(k)/V$ and $x_2=\sum_{k\neq 0} G_{33}(k)/V$ which represent atom and molecule pairs, respectively, as discussed in the supplement~\cite{supp}. Additionally $f_{12}=\sum_k G_{23}(k)/V$ characterizes the atom-molecule correlation function 
which plays a centrally important role. The equations of motion for the 2-point correlations can be compactly written as
\begin{eqnarray}
\label{5}
&&i\partial_t G_{mn}(k)=
\sum_\beta L_{m \beta}(k) G_{\beta n}(k)
 +L_{n \beta}(k) G_{m \beta }(k)
 \end{eqnarray}
where $L(k)$ represents the coupling between finite-momentum atoms and molecules~\cite{supp}. This dynamical scheme should be contrasted with the standard Hartree-Fock-Bogoliubov approach~\cite{Kokkelmans2002,Snyder2012} which omits these new
atom-molecule correlations such as $f_{12}$.

\begin{figure}[h]
\includegraphics[width=80mm,clip]{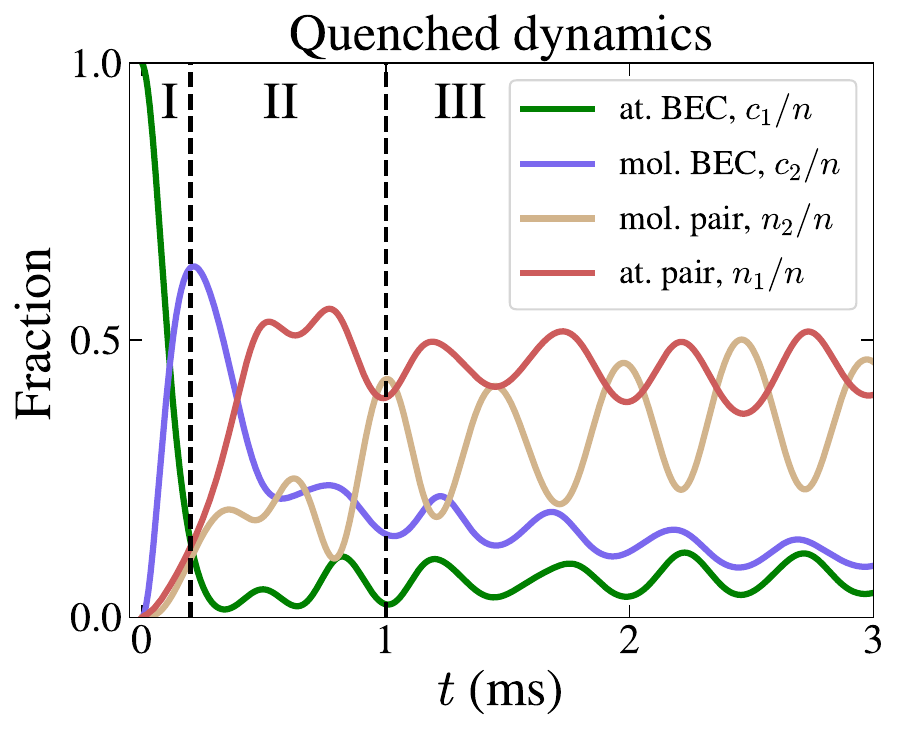}
\caption{ Population transfers and growth in the present theory.
When an atomic BEC is quenched to unitarity, a molecular BEC is the first to
be created. After
$t=0.2$
ms, both atomic and molecular BECs decay, generating a large fraction of non-condensed atoms and molecules. After 1 ms, all particles
exhibit oscillations with identical frequency. Only the non-condensed
molecules are out of phase as they control the flow of particles.}
\label{fig2}
\end{figure}

\begin{figure}[h]
    \centering
    \includegraphics[width=80mm,clip]
{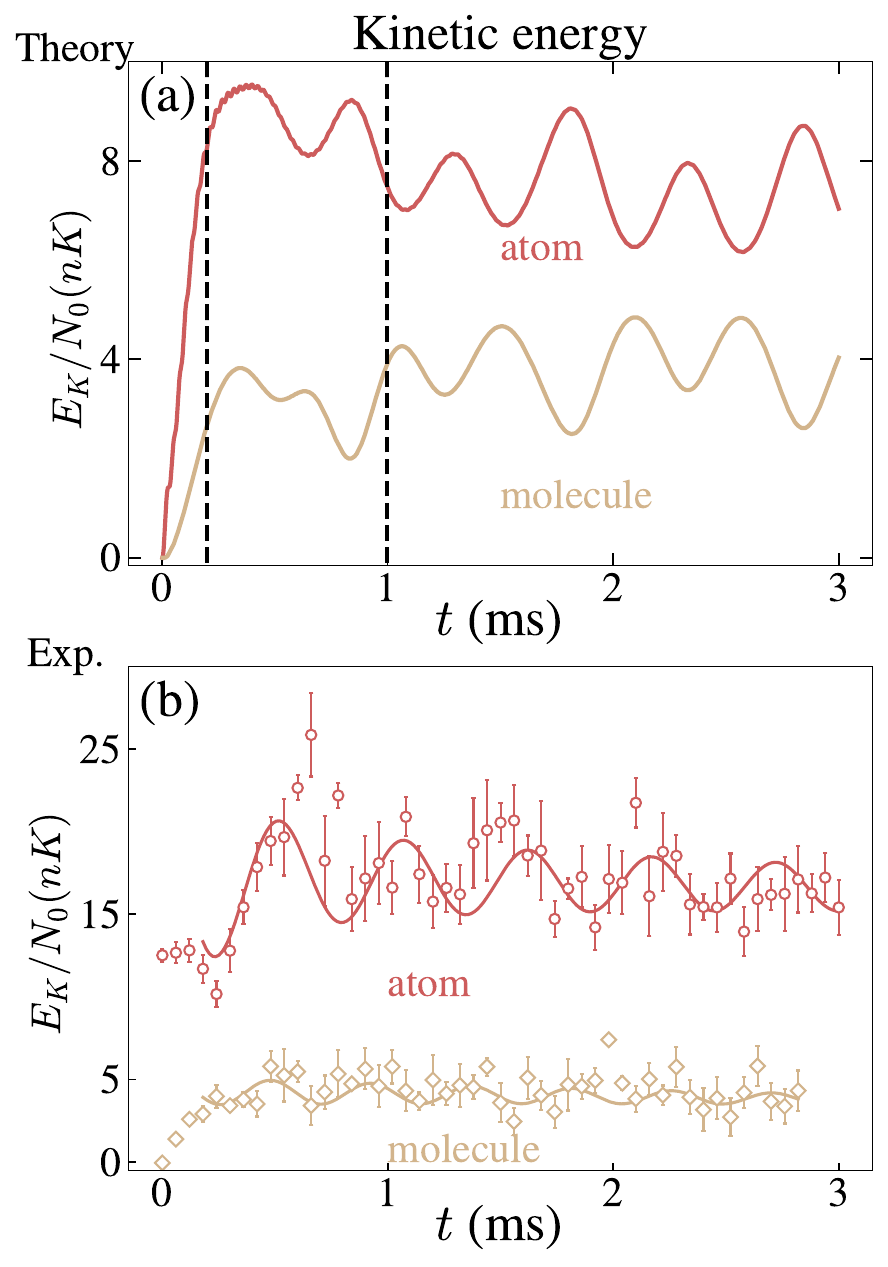}
    \caption{ Kinetic energy variation $E_K/N_0$  for atoms and molecules
which is associated with the population transfers and growth in Figure 2. Here $E^{\sigma}_K/N_0\equiv \int d^3k/(2\pi)^3 n_{\sigma}(k) (k_x^2+k_y^2)/(2m_{\sigma} n)$. 
Theory (top panel) and experiment (bottom panel).
In experiment (but not theory)
there is loss of particles
which is associated with a gain in kinetic energy, but the two
effects tend to at least partially compensate in the figure. Experiment and theory show
similar oscillatory behavior and
both show
a net increase of kinetic energy
which is of
the order of $5$ nK times $N_0$, where $N_0$ represents
the total number of particles at the start of the sweep. We focus here are the temperature variations
which are comparable, noting that there are differences in the absolute values presumably
because the experiments are at finite $T$, while the theory is at $T=0$. {  Thus, while panels (a) and (c) show some
quantitative differences, it is reasonable that that the occupation of
more and higher momentum states seen in experiment may reflect thermal
effects; the
experiments are performed at finite $T$, whereas theory is
at $T=0$.}
    }
    \label{fig3}
\end{figure}

\textit{Three Evolutionary Stages}
Using these dynamical equations we identify different stages of particle creation throughout the quench dynamics.
The quench of an atomic BEC to unitarity
is implemented by an abrupt change in the detuning $\nu_r\sim 0$  assuming
initially, at $t=0$, all particles reside in an atomic BEC.
As indicated in Fig.~\ref{fig2} 
in stage I, 
Feshbach couplings convert the initial atomic condensate into a coherent molecular BEC
along with a small number of atomic and molecular pairs having momentum, $\pm k$.
\footnote{
The growth of this molecular BEC is associated with instanton dynamics: $\xi_2(t)/\sqrt{n}=i \tanh(t/t_\alpha) $, where $t_\alpha=\hbar/\alpha \sqrt{2n}$~\cite{supp}}. 
In stage II, shortly
after its formation, the molecular BEC 
begins to partially decay and
more atomic pairs appear.
These latter in turn combine with an atom from the condensate to form an atom-molecule 
complex.
The combination of an atom-molecule complex and an atom from the condensate is
further converted into a pair of molecules. 
In stage III, with these
finite-momentum atomic 
and molecular populations now fully formed, the system enters a stage of steady-state coherent oscillations. 

What is particularly interesting  
about these oscillations is that, at unitarity, there is an in-phase relationship between all
quantities except for the non-condensed molecules. 
This phenomenon derives from a competition involving the three particle currents driven by the Feshbach coupling~\cite{supp}.
These currents represent the flow of particles between atoms
and molecules and condensates. Our central finding is that at resonance, 
once non-condensed molecules are created, 
they tend to control (through the equations of motion) the general dynamics of the particles.
As a consequence, the out-of-condensate
molecules are, interestingly, out-of-phase with the other three species~\cite{supp}.

\begin{figure*}
    \centering
\includegraphics[width=150mm]{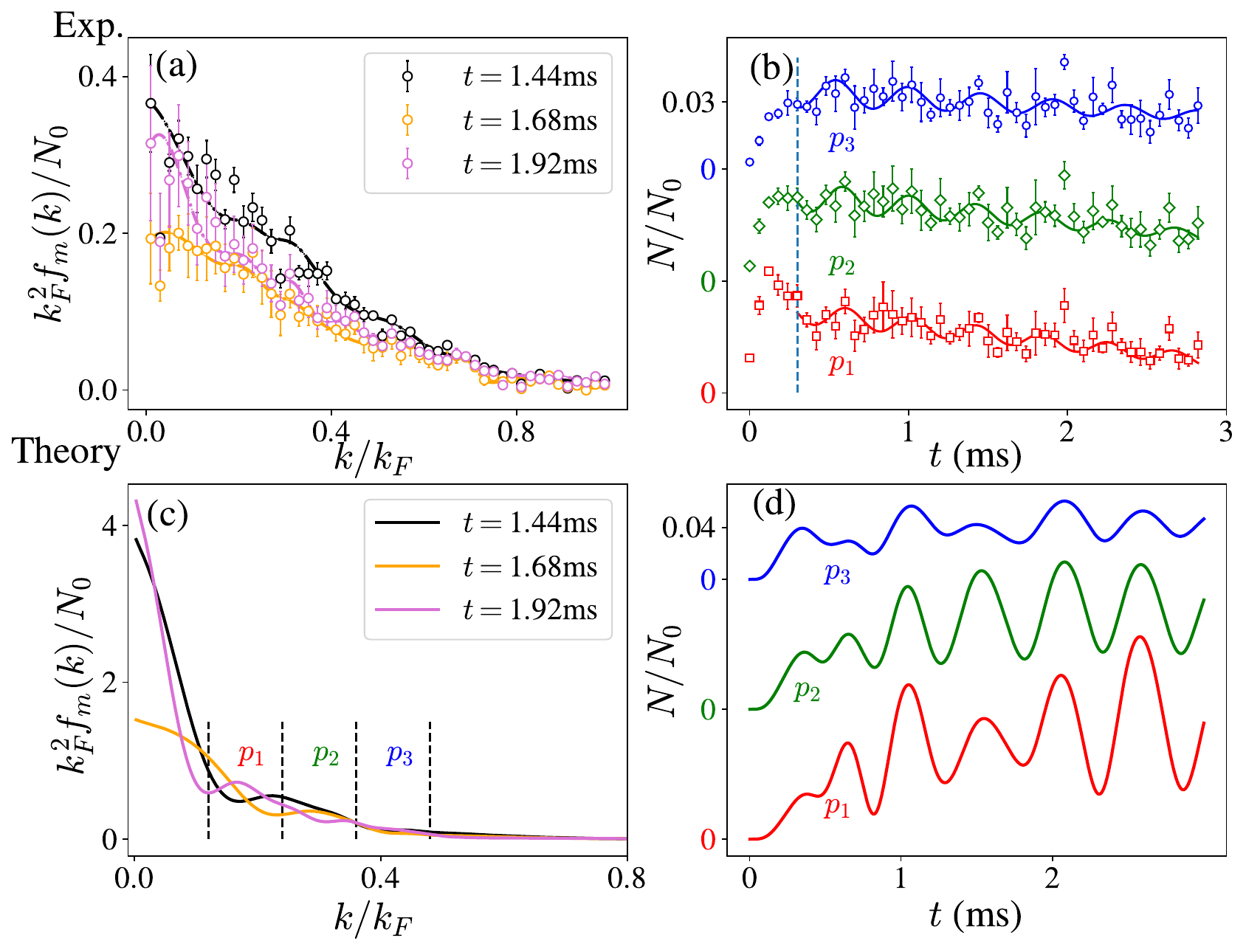}
    \caption{Theory-experiment comparison of the
dynamics of finite-momentum molecules where
(a) represents experimentally measured momentum distribution $f_m$ at 3 different times showing non-monotonicity-- the ordering of the curves from top to bottom corresponds to early, then late and then middle time. Here $f_m$ is the angular average momentum distribution of molecules such that $\int d^2k f_m=N_m$.
 (b) plots their number $vs$ time, $t$,
when summed over a momentum shell.  The summation region is indicated by $p_1/p_2/p_3$. Here $p_i=\int_{k_i\leq k<k_{i+1}}d^2k f_m(k)$ where $k_i=0.12k_F \times i $. Panels (c)
and (d) indicate the counterpart theory plots. We observe coherent oscillations in three different shells all with the same frequency of 2kHz in both theory and experiment. {  Comparing panels (a) and (c) suggests that the occupation of
more and higher momentum states seen in experiment may reflect thermal
effects:
the experiments are performed at finite $T$, whereas theory is
at $T=0$. }}
\label{fig4}
\end{figure*}

Most of the general observations we outline below (unless indicated otherwise) pertain to both theory and experiment 
\footnote{ Because the experimentally observed particle loss is not significant within the 3~$\mathrm{ms}$ time frame~\cite{Zhang2023}, it makes sense to consider these comparisons.}
and a comparison is presented in
Fig.~\ref{fig3}.
As is suggested by Figure 1, we see here that the newly generated out-of-condensate particles which carry kinetic energy are associated with
oscillatory dynamics. 
As the atomic condensate decays, a portion of its interaction energy is 
transferred to non-condensed particles.
This leads to an increase of the kinetic energy in both atoms and molecules immediately after the quench at $t<0.5~\mathrm{ms}$. We can quantify this
increase, $E^\sigma_K/N_0$, which is of the order of
10~$\mathrm{nK}$ for the atoms and  5~$\mathrm{nK}$ for the molecules, as found in both theory
and experiment.

After the initial growth period,
the oscillations persist for a period of around 2~$\mathrm{ms}$
notably, without any sign of thermalization. 
This oscillatory frequency  
is around 2 kHz, which, importantly, coincides with that of the population oscillations (see Fig.~\ref{fig2})~\cite{Zhang2023}.) 
Indeed, this single oscillation pattern observed in all
collective observables reflects the fact that the same underlying microscopic dynamical processes are
present.

{\it Momentum-resolved Distributions and Asymptotic Coherent Oscillations.} Of interest next is to clarify
the microscopic mechanism 
responsible for the generation of finite-momentum particles, after the initial
formation of the molecular BEC.
This is associated with the Feshbach resonant
interaction and 
is contained in the properties of the matrix $L(k)$ in Eq.~\ref{5}. The eigenvalues of $L$ determine the speed of growth of
these out-of-condensate particles while the eigenvectors determine the fraction of atoms and molecules involved
in the generation process\cite{PhysRevA.109.013316}.
We consider the eigenvalues of $L(k)$,
\begin{equation}
\lambda_\pm\simeq\sqrt{ F_1(k)\pm \sqrt{F^2_1(k)-4F_2(k)}    }.
\end{equation}
Here $F_1/F_2$ are given by $F_1(k)=\epsilon^2_1(k)+\epsilon^2_2(k)+8\alpha^2|\xi_1|^2-4\alpha^2|\xi_2|^2$ and  $F_2(k)=  (\epsilon_1(k) \epsilon_2(k)-4\alpha^2|\xi_1|^2 )^2- 4\alpha^2 |\xi_2|^2 \epsilon^2_2(k)$. 
This depends on dispersions given by $\epsilon_\sigma(k)=k^2/2m_\sigma+\nu_r \delta_{2,\sigma}$. 
One may observe that there exist two circumstances which lead
to an imaginary $\lambda$ 
and a consequent instability of condensates: either $F_1(k)<0$ which is associated
with the processes in stage I, or $F_2(k)<0$ which is mainly associated with 
those processes in stage II. These unstable outcomes are {\it unavoidable} when particles mostly reside in condensates. Indeed, once 
a molecular condensate is present ( $\xi_2\neq 0$), one can always find a value of $k$
such that $\epsilon_1 \epsilon_2-4\alpha^2|\xi_1|^2\simeq 0$ 
leading to an imaginary value for $\lambda_-$. Additionally, for $|\xi_2|\gg |\xi_1|$, the instability occurs at $\epsilon^2_1(k)+\epsilon^2_2(k)<4\alpha^2 |\xi_2|^2 $. In this way, finite-momentum particles are exponentially generated. 

All of this
can be contrasted with the larger $k$ regime where the eigenvalues are
real and 
there is consequently no significant occupation of those higher momentum states, having $k$ of the order
of, say, $k_{\mathrm{F}}$.
Here the kinetic oscillations rather than the exponential
growth dominates. 
Indeed, the momentum distributions we observe are consistent with the above analysis, where particles 
are seen to be concentrated in states with low momentum up to $\approx k_F/2$ where $k_F\equiv (6\pi^2n )^{1/3}$.
We stress that, in contrast to the saturation effects seen in the broad Feshbach resonance\cite{Makotyn2014,Hadzibabic}, the momentum distributions in the narrow
resonance case oscillate during the later time dynamics, as seen in Fig.~\ref{fig4}a,c.

To confirm that a single oscillation frequency is present in the experiments, as suggested by theory,
we determine the
molecular population in a shell including different groupings of momenta $p_i$,
as in Fig.~\ref{fig4}. 
Three such subsets were considered and the behavior in all is
found to be rather similar with all showing the characteristic frequency of
$2$ kHz.
This shared common frequency between finite-momentum atoms and molecules and their condensates should be understood
to derive from Feshbach coupling~\cite{supp}.  
We note parenthetically that the amplitude of the oscillations
is slightly greater in theory than experiment, which we attribute to finite temperature effects in our experiment that suppress the coherence of atoms and molecules.

A final challenge we must address is to understand 
the microscopic origin of the ubiquitous 2kHz oscillation frequency. 
Notably,
a frequency of this order is a large energy scale compared to other energy scales in the problem
including, for example, the self-interaction energy terms deriving from $g_1$ and $g_2$
which are of the order of $E_g/h\equiv g_{1/2} n \simeq  0.25$kHz.
To determine where this large frequency originates requires evaluating the separation of the energy levels between
atomic and molecular condensates. 
We are able to identify the main contribution to this energy gap as coming from the
new atom-molecule interspecies correlation $f_{12}$. This provides an additional self-energy correction to the atomic condensate $\mu_1 = -2 \alpha \text{Re} \,  f_{12}/\xi_1$, arising from the Feshbach coupling
$\sim \alpha  \hat{\psi}^\dagger_1(x)  \hat{\psi}^\dagger_1(x) \hat{\psi}_2(x) $ (see Eq.~\ref{4}).
Similarly, one finds the self-energy correction to the molecular condensate, $\mu_2=-\alpha \text{Re} \, x_1/\xi_2$. They together determine the splitting of the atomic and molecular condensate
levels: $\Delta \mu = 2 \mu_1 - \mu_2$. It is notable that without this correction to $\mu_1$,
the characteristic oscillation frequency 
would be around $0.9\text{ kHz}$\cite{wang2023stability}. But with these corrections, we find that
$\Delta \mu$ is, indeed, around $2$ kHz, bringing theory and experiment
into reasonable agreement~\cite{supp}.

{\it Conclusion.} 
We have emphasized throughout that the coherent oscillations we observe for the narrow resonance case
are universal as they are only determined by a {\it single} energy scale $\alpha \sqrt{n}$.
This leads to a different paradigm for the universal dynamics of the unitary Bose gas.
This is associated with a many body (Feshbach) coupling
and is to be contrasted with
the universality 
previously discussed in the literature
~\cite{Makotyn2014,Hadzibabic} and associated with the Fermi energy.
As in this early literature, arriving at this universality requires us to focus
on characterizing the dynamical behavior
of non-condensed particles which appear when
an atomic
condensate is swept to unitarity. Here in addition to atoms there are out-of-condensate molecules,
and once formed,
all populations reach a steady oscillatory state;
all finite-momentum particles
are seen to have the same intrinsic oscillation frequency
which is the same as that of the condensate~\cite{wang2023stability}.

In this context, we have presented a reasonably successful comparison between theory and experiment.
Because of
these quantitative correspondences we are able to glean
support for a rather new and different dynamical theory. This
is to be contrasted with the more conventional Hartree-Fock-Bogoliubov
approach ~\cite{Kokkelmans2002,Snyder2012}. Importantly, this dynamical machinery and its rich
phenomenology should be relevant to the 
general class of
bosonic superfluids associated with narrow Feshbach resonances.

{\it Acknowledgement} This work is supported by the National Science Foundation under Grant No. PHY1511696 and PHY-2103542, and by the Air Force
Office of Scientific Research under award number FA9550-21-1-0447. We thank Cheng Chin for very helpful discussions. Z.Z.
acknowledges the Bloch Postdoctoral Fellowship. S.N. acknowledges support from the Takenaka Scholarship Foundation. Z.W. is supported by the Innovation Program for Quantum Science and Technology (Grant No. 2021ZD0301904). We also
acknowledge the University of Chicago's Research Computing Center for their support of this work.

\bibliography{ref}
\end{document}


\title{
       Supplemental Material for  
       Universal coherent atom-molecule oscillations in the dynamics
of the unitary Bose gas near a narrow Feshbach resonance} 

\author{Ke Wang}
\affiliation{Department of Physics and James Franck Institute, University of Chicago, Chicago, Illinois 60637, USA}
\affiliation{Kadanoff Center for Theoretical Physics, University of Chicago, Chicago, Illinois 60637, USA}
  \author{Zhendong Zhang}
  \affiliation{E. L. Ginzton Laboratory and Department of Applied Physics, Stanford University, Stanford, CA 94305, USA}
\author{Shu Nagata}
\affiliation{Department of Physics and James Franck Institute, University of Chicago, Chicago, Illinois 60637, USA}
   \author{Zhiqiang Wang}
\affiliation{Department of Physics and James Franck Institute, University of Chicago, Chicago, Illinois 60637, USA}
\affiliation{Hefei National Research Center for Physical Sciences at the Microscale and School of Physical Sciences, University of Science and Technology of China, Hefei, Anhui 230026, China}

\affiliation{Shanghai Research Center for Quantum Science and CAS Center for Excellence in Quantum Information and Quantum Physics, University of Science and Technology of China, Shanghai 201315, China}

\affiliation{Hefei National Laboratory, University of Science and Technology of China, Hefei 230088, China
}
     \author{K. Levin}
     \affiliation{Department of Physics and James Franck Institute, University of Chicago, Chicago, Illinois 60637, USA}
     \maketitle

\setcounter{equation}{0}
\setcounter{figure}{0}
\setcounter{table}{0}
\setcounter{page}{1}
\makeatletter
\renewcommand{\thesection}{S\arabic{section}}
\renewcommand{\theequation}{S\arabic{equation}}
\renewcommand{\thefigure}{S\arabic{figure}}
\renewcommand{\bibnumfmt}[1]{[S#1]}

\section{Experimental background}

\subsection{ Typical experimental parameters and characterization of the momentum distribution}

For $^{133}$Cs the resonance at $B=19.849\,\text{G}$ has a very small width of $\Delta=8.3 \,\text{mG}$
. The experimental parameters are: the relative magnetic moment $\delta\mu=2\pi\hbar\times0.76 \,\text{MHz}/\text{G}$, the atomic s-wave scattering length $a_1=163a_0$ where $a_0$ is Bohr radius.  The molecular $s$-wave scattering length is $a_{2}=220a_0$. The mean BEC density is $n=3\times 10^{13} \text{cm}^{-3}$. {  In this way the coupling constants used in the simulations can be obtained and they are given by $g_1=3.28 E_F/k_F^3$, $g_2\simeq 2.43 E_F/k_F^3$ and $\alpha\simeq 1.53 E_F/k_F^{3/2}$, when the cut-off is chosen to be $a_1 \Lambda=\pi/10$. Here $E_F=\hbar^2 k_F^2/2m_1$.}

We extract experimentally from time of flight (TOF) measurements
the particle number density distribution in a two-dimensional space given by $n^{\text{TOF}}_{a/m}(x,y)$ 
which represents the atom and molecule particle number per pixel. Relating momentum and position variables $\hbar {\bf k}=m\omega {\bf x}$,
we find that
\begin{equation}
\label{1}
    {\bf k}_{a}=0.193 (\mu m^{-2}) \times {\bf x}_a,\quad   {\bf k}_{m}=0.387 (\mu m^{-2}) \times   {\bf x}_m.
\end{equation}
Here ${\bf k}_{a/m}$ represents the momentum of atoms or molecules while ${\bf x}_{a/m}$ represents
their positions.
In this way one can use Eq.~\ref{1} to obtain a particle number distribution in momentum space,
\begin{equation}
N_{a/m}(k_x,k_y,t)=n^{\text{TOF}}_{a/m}(x,y,t) \times S_{\text{pixel}}=n^{\text{TOF}}_{a/m}(x,y,t)\times (0.605 \mu m)^2
\end{equation}
Here we use the area of a single pixel $S_{\text{pixel}}=0.605^2 \mu m^2$. The mean BEC density is $n=3\times 10^{13} \text{cm}^{-3}$, which defines a momentum scale,
\begin{equation}
k_F=(6\pi^2 n)^{1/3}\simeq 12.1 \mu \text{m}^{-1}.
\end{equation}
To visualize the 2D data as a 1D curve, we perform the angular average below to obtain a momentum distribution,
\begin{equation}
f_{a/m}(k_i,t):=\frac{1}{2\pi (k_i+\Delta/2) \, \Delta  } \sum_{ k_i\leq \sqrt{k_x^2+k_y^2}< k_i+\Delta  } N_{a/m}(k_x,k_y,t),\quad k_i=i\Delta.
\label{4}
\end{equation}
Here $f$ has the dimensions of area and $\Delta=k_F/50 $. This allows us to obtain
the momentum distribution as given in the main-text with the normalization
\begin{equation} 
\int d^2k f_{a/m}(k,t)  =N_{a/m} 
\end{equation}
 
\begin{figure}[h]
    \centering    \includegraphics[width=0.5\linewidth]{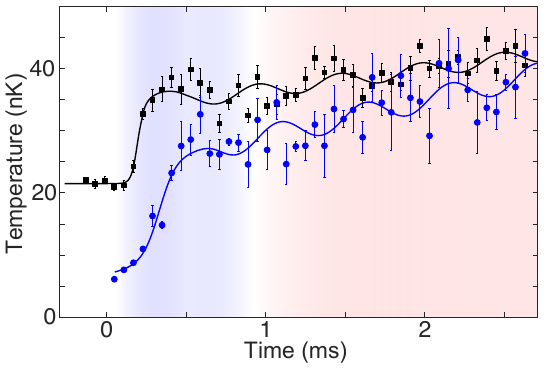}
    \caption{Effective temperature of atoms (black points) and molecules (blue points) from Ref.~[17]. Data points at $t>0.5$ ms are fit using the function $y(t) = A\sin{(2\pi ft+\phi)+Bt+C}$. The fit determines the oscillation frequency $f$ to be 2.0(2) kHz for atoms and 1.9(2) kHz for molecules. The parts of the curves at $t\leq 0.5$ ms are guides to the eye.}
    \label{fig:Refit_T}
\end{figure}

\subsection{Alternative fit of the effective temperature data in Ref.~[17].}
In contrast to an earlier empirical fit to the effective temperature, in Fig 1 we use a sinusoidal function plus a linear increasing offset $y(t) = A\sin{(2\pi ft+\phi)+Bt+C}$ to fit the coefficients (A,B,C) based on the error bars
in the data. In this way
we can replot the atomic and molecular temperature data presented in Fig.3b of Ref.~[17]. We choose to fit the data points at $t>0.5$ ms when they start to oscillate. The oscillation frequencies are determined to be 2.0(2) kHz for atoms and 1.9(2) kHz for molecules, which are consistent with the universal 2 kHz oscillation frequency seen in other observables shown in the main text.


\vskip10mm 

\section{ Theoretical Analysis}

{ \subsection{Two-dimensional momentum distribution}}
We theoretically compute the particle number distribution in momentum space, $n(k)$. When integrated over
momentum this yields the density in each channel, $n=\int\frac{d^3k }{(2\pi)^3} n(k).$
Of interest is the integral along the $z$-axis representing a column integrated
two dimensional distribution. We define
\begin{eqnarray}
F_{a/m}(k_\perp)= \int\frac{dk_z}{2\pi} n_{a/m}\left(\sqrt{k_\perp^2+k_z^2}\right)
\end{eqnarray}
Imposing a normalization condition $(2\pi)^{-2}\int d^2k_\perp F_{a/m}=N_{a/m}/V$ yields the following relation between $F_{a/m}$ and the variable $f_{a/m}$, 
\begin{eqnarray}
\frac{V}{(2\pi)^2} F_{a/m}(k)=f_{a/m}(k)\iff \frac{N_0/n}{(2\pi)^2} F_{a/m}(k)=f_{a/m}(k),
\end{eqnarray}
which allows us to relate the theory to experimental data $f_{a/m}$. Here $N_0$ is the initial particle number. The dimensionless quantity plot in Fig.~3c of the main-text is
based on the relation: 
\begin{eqnarray}
\frac{ k_F^2 f_{a/m}}{N_0}=(2\pi)^2\frac{k_F^2}{n}  F_{a/m}(k).
\end{eqnarray}

\subsection{Derivation of the Dynamical Equations}
In this section, we discuss the 2-channel model and the associated dynamical equations. We start from the effective field theory 
of the Hamiltonian
\begin{eqnarray}
\label{5}
 H=  \sum_\sigma \int d^d x \hat{\psi}^\dagger_\sigma(x) \left( -\frac{\nabla^2}{2m_\sigma}+\nu_\sigma \right)  \hat{\psi}_\sigma(x)+   \int d^d x   \left(\sum_\sigma \frac{g_\sigma}{2} |\hat \psi_\sigma(x)|^4  
    - \alpha  \hat{\psi}^\dagger_1(x)  \hat{\psi}^\dagger_1(x) \hat{\psi}_2(x) -h.c.  \right).
\end{eqnarray}
Here $\sigma=1,2$ refers to the open channel atoms and closed-channel molecules, $g_\sigma$ represents the background $s$-wave interaction in the open and closed channels and $\alpha$ represents
the Feshbach coupling strength between two channels. Here $\nu_1=0$ and $\nu_2-\nu_1$ 
is related to the binding energy of molecules. This model allows us to
address the kinematics of the open channel atoms and closed channel molecules, and the couplings between them. 
We note that with the appropriate parameters we are able to reproduce the resonance width and the 2-body scattering length of atoms. 

To address the quantum dynamics we relate the fundamental dynamical variables in our model to the equal-time correlation functions of field operators. Here the one-point correlation function represents the condensate wave-function,
\begin{eqnarray} 
\xi_\sigma=\frac{1}{\sqrt{V}} \langle \hat{\psi}_\sigma(k=0 )\rangle,\quad \sigma=1,2. 
\end{eqnarray}
With this definition we deduce the {\it exact} equation of motion for condensates,
\begin{eqnarray}
i  \frac{d}{dt} \xi_{1}  &=& 2g_1  n_1\xi_{1}  +\left(g_1 x'_1-2\alpha \xi_2  \right)\xi^*_1- {2\alpha}f_{12}  +g_1 f_{111} \nonumber \\
i  \frac{d}{dt} \xi_{2}  &=& \left( \nu+2g_2  n_2\right)\xi_{2} + g_2 x'_2 \xi^*_{2}-2\alpha x'_1  + g_2 f_{222} 
\end{eqnarray}
where we denote
\begin{eqnarray}  
&&f_{12}=\int d^3q/(2\pi)^3\langle \psi'^\dagger_1(q) \psi'_2(q)\rangle,\quad  f_{\sigma\sigma\sigma}=\int d^3q_1 d^3q_2 /(2\pi)^6 \langle \psi'^\dagger_\sigma(q_1)  \hat \psi'_\sigma(q_2) \hat \psi'_\sigma(q_1-q_2) \rangle . \\
&& n_\sigma=\int d^3q/(2\pi)^3\langle \psi'^\dagger_\sigma(q) \psi'_\sigma(q)\rangle,\quad x'_\sigma=\xi^2_\sigma+\int d^3q/(2\pi)^3\langle \psi'_\sigma(q) \psi'_\sigma(-q)\rangle.
\end{eqnarray}
Here $n_\sigma$ represents the number of particles in finite momentum states.

The correlation functions are given in terms of $\hat \psi_\sigma'(k)=\hat \psi_\sigma(k)-\delta_{k,0} \xi_\sigma$ 
where we have introduced a 4-vector field operator, $\Psi=(\psi'_1, \psi'^\dagger_1, \psi'_2, \psi'^\dagger_2 )$.  One may define a general $n$-point correlation function below,
\begin{eqnarray} 
G_{i_1,i_2,...,i_n}(k_1, k_2, ..., k_n)=\langle \hat\Psi_{i_1}(k_1) \hat\Psi_{i_2}(k_2) ... \hat\Psi_{i_n}(k_n)\rangle. 
\end{eqnarray}

Next we derive the dynamics of these correlation functions using the time evolution operator, 
$\hat U(t)=\exp(-i \hat{H} t)$ based on the Hamiltonian in Eq.~\ref{5}. This yields the following equations of motion:
\begin{eqnarray}
 i\frac{d}{dt} \hat \Psi_{i}(k) =L'_{i \beta}(k) \hat\Psi_\beta(k) + \sum_{q}Q_{i \beta \gamma } \hat\Psi_\beta(q) \hat \Psi_\gamma(k-q)+\sum_{q_1,q_2} C_{i \beta \gamma \delta }\hat \Psi_\beta(q_1) \hat\Psi_\gamma (q_2)\hat\Psi_\delta(k-q_1-q_2). \nonumber
\end{eqnarray}
Here $L'$ is 2 point vertex related to the quadratic part of Hamiltonian, $Q$ is the 3-point vertex which includes the Feshbach coupling term while $C$ is a 4-point vertex including background scatterings. From the bosonic commutators, one can readily derive the expression of $L'$,
\begin{eqnarray}
 L'=\begin{pmatrix}
\epsilon'_1(k) & g_1\xi_1^2-2\alpha \xi_2 & -2\alpha \xi^*_1    & 0 \\
-(g_1 \xi_1^2-2\alpha \xi_2)^* & -\epsilon'_1(-k) & 0 & 2\alpha\xi_1 \\
-2\alpha \xi_1 & 0 &  \epsilon'_2(k) & g_2 \xi^2_2 \\
0 & 2\alpha \xi^*_1 & -(g_2 \xi^2_2)^* & -\epsilon'_2(-k) 
\end{pmatrix},\quad \epsilon'_\sigma(k)= k^2/2m_i+\nu_r \delta_{2,i}+2g_i |\xi_i|^2.
\end{eqnarray}
Only a few values of $Q_{\alpha \beta \gamma }$ 
are non-vanishing. These are given by
\begin{eqnarray}
Q_{121}=\frac{2g_1\xi_1}{\sqrt{V} }, \,  \, Q_{111}=\frac{ g_1\xi^*_1}{\sqrt{V} }, \,  \, Q_{123}=-\frac{2\alpha}{\sqrt{V}};   \,\, Q_{221}=-\frac{2g_1\xi^*_1}{\sqrt{V} }, \,  \, Q_{222}=-\frac{ g_1\xi^*_1}{\sqrt{V} }, \,  \, Q_{214}=\frac{2\alpha}{\sqrt{V}}; \nonumber \\
Q_{343}=\frac{2g_2\xi_2}{\sqrt{V} }, \,  \, Q_{333}=\frac{ g_2\xi^*_2}{\sqrt{V} }, \,  \, Q_{311}=-\frac{\alpha}{\sqrt{V}};  \,\,  
Q_{443}=-\frac{2g_2\xi^*_2}{\sqrt{V} }, \,  \, Q_{444}=-\frac{ g_2\xi_2}{\sqrt{V} }, \,  \, Q_{422}=\frac{\alpha}{\sqrt{V}};
\end{eqnarray}
Similarly, only a few values for $C$ are non-zero,
\begin{eqnarray}
C_{1211}=\frac{g_1}{V},\, \, C_{2221}=-\frac{g_1}{V}, \, \,C_{3433}= \frac{g_2}{V},  \, \,  C_{4443}=-\frac{g_2}{V}.
\end{eqnarray}

This leads to the following heirarchy of equations,
\begin{eqnarray}
\label{11}
i\frac{d}{dt}G_{i_1, i_2, ..., i_n}(k_1,..., k_n)=\sum_{j,\beta}L'_{i_j \beta}(k_j) G_{i_1, ...,i_{j-1} , \beta, i_{j+1},...i_n}(k_1,..., k_n) \nonumber \\
+\sum_{j,\beta \gamma }Q_{i_j \beta \gamma} \sum_{q} G_{i_1, ...,i_{j-1} , \beta, \gamma, i_{j+1},...i_n}(k_1,...,k_{j-1},q,k_j-q, k_{j+1},... k_n) \nonumber \\
+\sum_{j,\beta \gamma \delta }C_{i_j \beta \gamma \delta } \sum_{q_1,q_2} G_{i_1, ...,i_{j-1} , \beta, \gamma, \delta, i_{j+1},...i_n}(k_1,...,k_{j-1},q_1,q_2,k_j-q_1-q_2, k_{j+1},... k_n)
\end{eqnarray}
which relates $n$-point correlations with $n \,(L), n+1 \,(Q), n+2 \,(C)$ correlations.  

Now consider a quench of a pure atomic condensate to unitarity. It is useful
to consider the way in which the dynamics reflects a sequence of {\it propagation of information} in the correlation functions,
\begin{eqnarray}
\label{12}
\xi_1\longrightarrow \xi_2 \longrightarrow G_{\alpha \beta} \longrightarrow G_{\alpha \beta \gamma} \longrightarrow ....
\end{eqnarray}
Correspondingly, we can consider a sequence of points on the time axis,
\begin{eqnarray}
t=0 \longrightarrow t_1 \longrightarrow t_2 \longrightarrow t_3  \longrightarrow ...
\end{eqnarray}
Because of the small size of the coupling constant $\alpha$, 
one may have well-separated dynamical regimes 
\begin{eqnarray}
\text{I}=[0,t_1],\, \text{II}=[t_1,t_2],\,   \text{III}=[t_2,t_3], ...
\end{eqnarray}

We stress here the importance of the subscript $i$ in this discussion.
The $i$ associated with $t_i$ indicates the time interval where the $i$-point correlation is fully developed and becomes non-perturbative. For example, in the time interval associated with $t_2$ we
need only include 2-point correlation functions in modeling the dynamics.

We observe that with increasing time the atomic and molecular BEC's decay and the coherence of the dynamical system is significantly reduced after 
the time frame $t_2$.
Since the coherence greatly enhances the speed of reaction/collision events, we expect that the time scale of each stage 
monotonically increases such that $t_3-t_2\gg t_2-t_1\gg t_1$. Therefore the time required to fully develop the 3-point correlation is much longer than the $t_2$, which is the order of $1$ms.
Within the experimentally accessible time scale ($< 3$ms), one may conclude that the contribution of 3-point correlation to the dynamics is subleading compared to the 2-point correlation. As a consequence in modeling the dynamics
we set 3-point correlations to be zero and decompose 4-point correlations into products of 2-point correlation
functions. Note that 3-point correlations are indecomposable since the average of $\psi'$ is zero by definition. This then leads to a closed form for the dynamical equations (now consisting of
only one-point and two-point functions.)  

Within this approximation, one can derive a closed set of dynamical equations that include one- and two-point correlation functions. In this scheme, the two-point correlation function, \(L'\), receives corrections from the decomposition of four-point correlations. This leads to the expression for the dynamical matrix, \(L\), which appears in Eq.~3 of the main text,
 \begin{eqnarray}
 L=\begin{pmatrix}
\epsilon_1(k) & g_1x'_1-2\alpha \xi_2 & -2\alpha \xi^*_1    & 0 \\
-(g_1 x'_1-2\alpha \xi_2)^* & -\epsilon_1(-k) & 0 & 2\alpha\xi_1 \\
-2\alpha \xi_1 & 0 &  \epsilon_2(k) & g_2 x'_2 \\
0 & 2\alpha \xi^*_1 & -g_2 x'^{*}_2 & -\epsilon_2(-k) 
\end{pmatrix}
\end{eqnarray}
where $\epsilon_i(k)=k^2/2m_i+\nu_r \delta_{2,i}+2g_i n_i$.

\subsection{Different stages in the dynamical evolution}

After the quench of the atomic BEC, the next most immediate stage of the dynamical evolution involves
the creation of a molecular condensate. 
To describe this
we begin by considering the dynamics with only two condensates, $\xi_1$ and $\xi_2$. We set $g_1=g_2=0$ for simplicity.  
Since our focus is on unitarity, we set the detuning to zero.  This leads to the coupled equations below,
\begin{eqnarray}
i  \partial_t \xi_{1}  &=&  -2\alpha \xi_2   \xi^*_1 ,\quad
i  \partial_t \xi_{2} = 
 - \alpha \xi^2_{1}.
\end{eqnarray}
One may use the constraint from the conservation of particle-number $|\xi_1|^2+2|\xi_2|^2=n$ to eliminate $\xi_1$ and find a single equation for $\xi_2$,
\begin{eqnarray}
\partial^2_t \xi_{2}+4 \alpha^2  \xi_{2}(n-|\xi_2|^2)=0.
\end{eqnarray}
With the initial condition from the quenched dynamics $\xi_2(0)=0,\, \partial_t\xi_2(0)=i\alpha n$, 
the differential equation allows an instanton solution,
\begin{eqnarray}
  \xi_2=i \sqrt{n}\tanh\left(
 \alpha \sqrt{2n} t
\right).
\end{eqnarray}
This in turn sets a time scale for the growth of the molecular condensate, $t_\alpha=\hbar/\alpha \sqrt{n}$. Due to the narrow Feshbach resonance, this time is not particularly small and the time scale should be observable in
experiment as it is of the order of $0.2$ms. Such instanton phenomena are characterized by a sharp peak in the particle fraction of the molecular BEC as a function of time which can be seen in Fig.~2 of the maintext. 

\subsection{Determining the Oscillation Frequency}
\begin{figure}
    \centering
\includegraphics[scale=0.4]{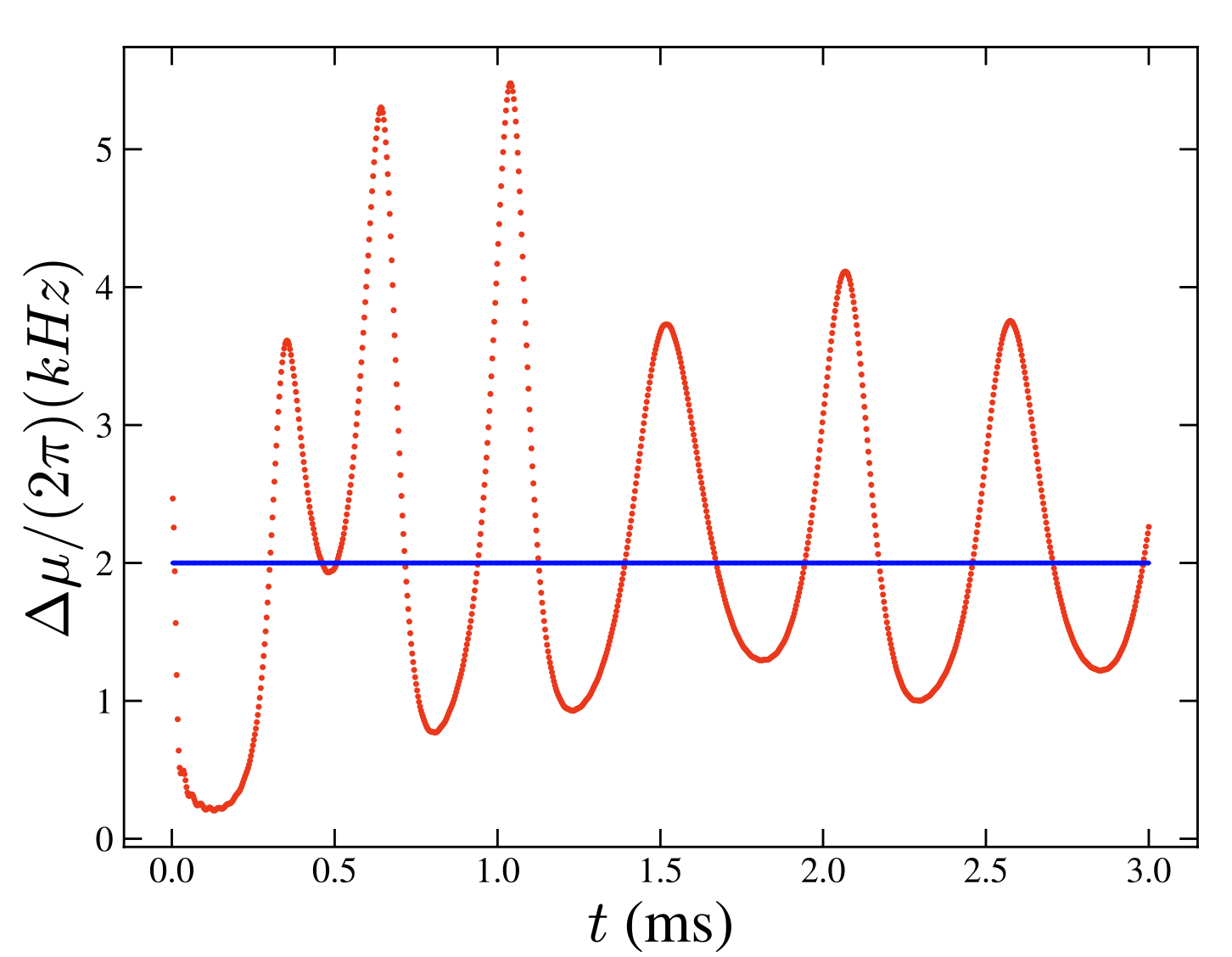}
    \caption{The figure plots $\Delta \mu$ (which is the time dependent energy level difference
between the two condensates), as the function of time.  Here $h=1$. Numerically we find that the average of $\Delta \mu$ in the late time is around $2.1$kHz.}
    \label{gap}
\end{figure}
In the main text, we note that a gap is dynamically generated between the atomic and molecular BEC. This gap originates from the energy level difference \(\Delta \mu=2\Delta \mu_1-\Delta \mu_2\) between the two condensates. Here, \(\Delta \mu_\sigma\) represents the self-energy correction. One may derive \(\Delta \mu_1\) from rewriting the condensate dynamics:
\begin{eqnarray}
  i  \frac{d}{dt} \xi_{1}  &=& 2g_1  n_1\xi_{1}  (0)+\left(g_1 x_1-2\alpha \xi_2  \right)\xi^*_1-2\alpha \xi_1 \left( \text{Re} \frac{f_{12}}{\xi_1}+ i  \text{Im} \frac{f_{12}}{\xi_1}\right),\quad f_{12}=\int \frac{d^3k}{(2\pi)^3} f_{12}(k).
\end{eqnarray}
Similarly, one can derive $\Delta \mu_2$. Therefore one can obtain:
\begin{equation}
     \Delta \mu_1 = -2 \alpha \text{Re} \, \xi^{-1}_1{ \sum_{k}\Big\langle \psi^\dagger_1(k) \psi_2(k) \Big\rangle},\quad \Delta \mu_2=-\alpha \text{Re} \, \xi^{-1}_2{ \sum_{k}\Big\langle \psi_1(k) \psi_1(-k) \Big\rangle}.
\end{equation}
These functions \(\Delta \mu\) are time-dependent and oscillate around a constant value. The physical energy gap should be determined by the time-average \(\langle\Delta \mu\rangle_t\); see Fig.~\ref{gap}.

The finite momentum particles 
will also oscillate
at the same frequency as the condensates. Because they have
a generally kinetic energy,
$ \epsilon_k/h \ll 2 \mathrm{kHz}$, this will not appreciably alter
the characteristic frequency.

\subsection{Coherent Particle Flow Currents}
There are four species of particles in the system. Let us denote them by  
\begin{eqnarray}
   && c_1,c_2: \text{This labels the particles in the atomic and molecular BECs}, \\
   && n_1, n_2: \text{This labels the finite momentum particles for atoms or molecules}. 
\end{eqnarray}
\begin{figure}
    \centering  \includegraphics[width=0.5\linewidth]{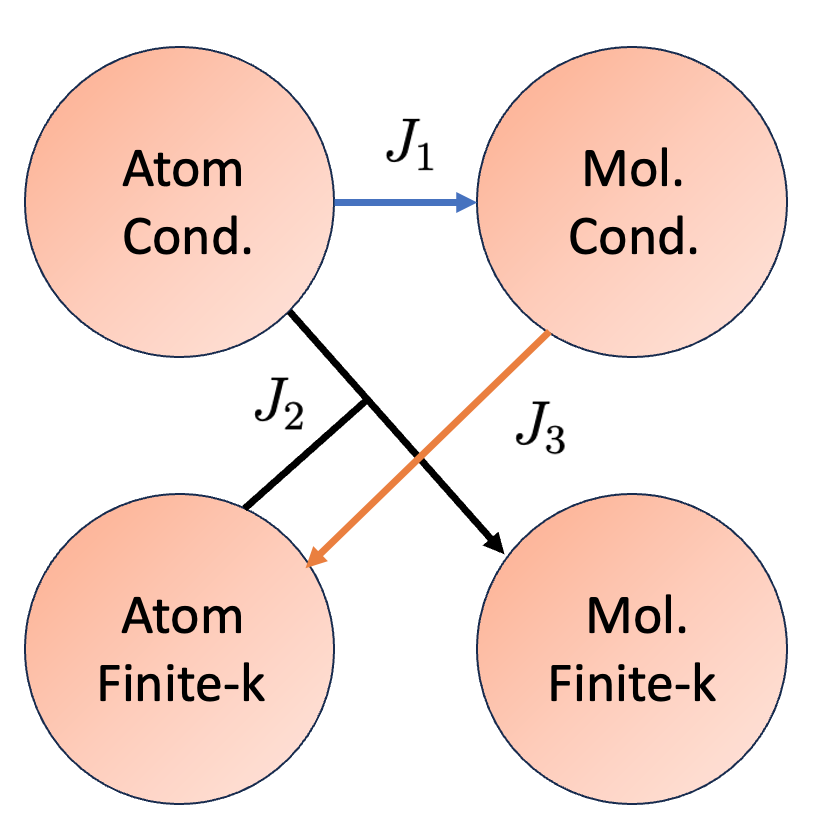}
    \caption{ There are three types of currents flow. (1) $J_1$: two cond. atoms form a molecule. (2). $J_2$: One cond. atom combine with one finite-$k$ atom, forming a finite momentum molecule. (3)
 $J_3$. One cond. Molecule dissociate into two finite-k atoms.
    }
    \label{conversion}
\end{figure}
Particles flow among all four species as plotted in Fig.~\ref{conversion}. The equations governing the
dynamics (and counted in terms of the number of atoms) are:
\begin{eqnarray}
&& \frac{d}{dt} c_1=-J_1-J_2,\quad \frac{d}{dt} c_2=J_1/2-J_3 ,\quad \frac{d}{dt} n_1=-J_2+2J_3,\quad \frac{d}{dt} n_2=J_2 .
\end{eqnarray}
The currents are controlled by the phases of condensates and of the various finite-momentum correlation
functions. This leads to expressions for the various currents:
\begin{eqnarray}
&& J_1=4\alpha c_1 \sqrt{c_2} \sin(\phi_2-2\phi_1),\quad J_2=4\alpha   |f_{12}| \sqrt{c_2} \sin(\theta_f-\phi_1),\quad J_3=2\alpha |x_1| \sqrt{c_2} \sin(\theta_x-\phi_2).
\end{eqnarray}
Here $\phi_1$ and $\phi_2$ are phases of atomic and molecular condensates, and $\theta_f$ and $\theta_x$ are phases of the complex numbers $f_{12}$ and $x_1$.
  
All finite momentum particles collectively determine these two phases. The net particle flow results
from a competition between these three different current which mostly consist of low $k$ states.
At resonance, we find that $|J_2|>|J_1|>2|J_3|$
and that $J_2$ has a different sign from $J_1$ and $J_3$ at least after $1$ms. It then follows that
the out-of-condensate 
molecules (represented by $n_2$) are out-of-phase with the other three species of particles.

\subsection{The generation of Atom-molecule complexes and molecule-molecule pairs}
In this section, we expand the discussion surrounding Eq.~3 in the main-text to illustrate how the atomic and molecular pairs are generated through the 2-point correlation functions involving the Feshbach coupling,
\begin{eqnarray}
  i \partial_t   n_1(k)&=&   -2\alpha \Big( 
  f_{12}(k)\xi_1^\dagger +  x^*_{1}(k) \xi_2-   - h.c.
\Big), \quad i \partial_t   n_{ 2}(k)=   -2\alpha \Big( 
 f^*_{12}(k)\xi_1 -h.c.
\Big)   \nonumber
  ,\\
  \left(i \partial_t+ \frac{k^2}{m} \right) x_1(k)&=& -2\alpha \left( 
 2 a_{12}(k)\xi^\dagger_1  + (2n_1(k)+1 ) \xi_2 
\right),\quad \left(i \partial_t +\frac{k^2}{2m } \right) x_{2}(k)=  -2\alpha \Big( 
  a_{21}(k)\xi_1  +\xi_1 a_{12}(k)\Big). \nonumber \\
  \left(i \partial_t+\frac{3k^2}{4m } -\nu \right) a_{12}(k)&=& -2\alpha \Big( 
  x_{2}(k)\xi^\dagger_1+\xi_2   f_{12}(k)   + x_{1}(k) \xi_1 
\Big),\,  \left( i \partial_t -\frac{k^2}{4m} -\nu\right) f_{12}(k)=  -2\alpha \Big( 
  n_{1}(k)\xi_1 -  n_{2}(k) \xi_1- a_{12}(k) \xi^\dagger_2 \Big).\nonumber
\end{eqnarray}
Here $a_{12}(k)=\langle \psi'_1(k) \psi'_2(-k)\rangle=a_{21}(-k)$. In a more concrete example, assume that one starts from a molecular condensate, and considers the
simultaneous dissociation of an molecular condensate ($\xi_2$) into atom pairs ($x_1$). Then this pair of atoms can combine with an atom to form an atom-molecule complex, ($a_{12}$), which can then combine with an atom and further generate a molecule pair ($x_2$).